\documentclass[aps,prd,floats,preprint,nofootinbib]{revtex4}
\usepackage{graphicx}
\usepackage{amsmath}
\usepackage{amsfonts}
\usepackage{amssymb}
\usepackage{multirow}
\begin{document}
\def\beq{\begin{equation}}
\def\eeq{\end{equation}}
\def\bea{\begin{eqnarray}}
\def\eea{\end{eqnarray}}
\def\ben{\begin{enumerate}}
\def\een{\end{enumerate}}
\def\ie{{\it i.e.}}
\def\etc{{\it etc.}}
\def\eg{{\it e.g.}}
\def\lsim{\mathrel{\raise.3ex\hbox{$<$\kern-.75em\lower1ex\hbox{$\sim$}}}}
\def\gsim{\mathrel{\raise.3ex\hbox{$>$\kern-.75em\lower1ex\hbox{$\sim$}}}}
\def\ifmath#1{\relax\ifmmode #1\else $#1$\fi}
\def\half{\ifmath{{\textstyle{1 \over 2}}}}
\def\threehalf{\ifmath{{\textstyle{3 \over 2}}}}
\def\quarter{\ifmath{{\textstyle{1 \over 4}}}}
\def\eigth{\ifmath{{\textstyle{1\over 8}}}}
\def\sixth{\ifmath{{\textstyle{1 \over 6}}}}
\def\third{\ifmath{{\textstyle{1 \over 3}}}}
\def\twothirds{{\textstyle{2 \over 3}}}
\def\fivethirds{{\textstyle{5 \over 3}}}
\def\fourth{\ifmath{{\textstyle{1\over 4}}}}
\def\chitil{\wt\chi}
\def\fbi{~{\mbox{fb}^{-1}}}
\def\fb{~{\mbox{fb}}}
\def\br{BR}
\def\gev{~{\mbox{GeV}}}
\def\calm{\mathcal{M}}
\def\mll{m_{\ell^+\ell^-}}
\def\tanb{\tan\beta}

\def\wtil{\widetilde}
\def\cnone{\wt\chi^0_1}
\def\cnonestar{\wt\chi_1^{0\star}}
\def\cntwo{\wt\chi^0_2}
\def\cnthree{\wt\chi^0_3}
\def\cnfour{\wt\chi^0_4}
\def\snu{\wt\nu}
\def\snul{\wt\nu_L}
\def\msnul{m_{\snul}}
\def\se{\wt e}
\def\smu{\wt\mu}
\def\snu{\wt\nu}
\def\snul{\wt\nu_L}
\def\msnul{m_{\snul}}

\def\snue{\wt\nu_e}
\def\snuel{\wt\nu_{e\,L}}
\def\msnuel{m_{\snul}}

\def\snubar{\ov{\snu}}
\def\msnu{m_{\snu}}

\def\snue{\wt\nu_e}
\def\snuel{\wt\nu_{e\,L}}
\def\msnuel{m_{\snul}}

\def\snubar{\ov{\snu}}
\def\msnu{m_{\snu}}
\def\mcnone{m_{\cnone}}
\def\mcntwo{m_{\cntwo}}
\def\mcnthree{m_{\cnthree}}
\def\mcnfour{m_{\cnfour}}
\def\wt{\widetilde}
\def\anti{\overline}
\def\wh{\widehat}
\def\cpone{\wt \chi^+_1}
\def\cmone{\wt \chi^-_1}
\def\cpmone{\wt \chi^{\pm}_1}
\def\mcpone{m_{\cpone}}
\def\mcpmone{m_{\cpmone}}

\def\staur{\wt \tau_R}
\def\staul{\wt \tau_L}
\def\stau{\wt \tau}
\def\mstaur{m_{\staur}}
\def\stauone{\wt \tau_1}
\def\mstauone{m_{\stauone}}

\def\gl{\wt g}
\def\mgl{m_{\gl}}
\def\stl{{\wt t_L}}
\def\str{{\wt t_R}}
\def\mstl{m_{\stl}}
\def\mstr{m_{\str}}
\def\sbl{{\wt b_L}}
\def\sbr{{\wt b_R}}
\def\msbl{m_{\sbl}}
\def\msbr{m_{\sbr}}
\def\sbot{\wt b}
\def\msbot{m_{\sbot}}
\def\sq{\wt q}
\def\sqbar{\ov{\sq}}
\def\msq{m_{\sq}}
\def\slep{\wt \ell}
\def\slepbar{\ov{\slep}}
\def\mslep{m_{\slep}}
\def\slepl{\wt \ell_L}
\def\mslepl{m_{\slepl}}
\def\slepr{\wt \ell_R}
\def\mslepr{m_{\slepr}}
\def\jet{{\rm jet}}
\def\filt{{\rm filt}}
\def\cut{{\rm cut}}
\def\sub{{\rm sub}}
\def\sig{\mbox{sig}}
\def\ch{\text{ch}}
\def\cut{\text{cut}}
\def\nsub{ n_{\text{sub}}}
\def\track{\text{track}}
\def\nch{N_{\text{ch}}}
\def\mfilt{m_{\text{filt}}}
\def\anch{\langle\nch\rangle}
\def\CC{{C\nolinebreak[4]\hspace{-.05em}\raisebox{.4ex}{\tiny\bf ++}}}
\newcommand{ \slashchar }[1]{\setbox0=\hbox{$#1$}   
   \dimen0=\wd0                                     
   \setbox1=\hbox{/} \dimen1=\wd1                   
   \ifdim\dimen0>\dimen1                            
      \rlap{\hbox to \dimen0{\hfil/\hfil}}          
      #1                                            
   \else                                            
      \rlap{\hbox to \dimen1{\hfil$#1$\hfil}}       
      /                                             
   \fi}
\widowpenalty=1000
\clubpenalty=1000
\vspace*{3cm}
\title{Tracking the Identities of Boosted Particles}

\author{Zhenyu Han}

\affiliation{ \small \sl Center for the Fundamental Laws of Nature\\ Harvard University,\\ Cambridge, MA 02138, USA}

\def\thesection{\arabic{section}} 
\def\thetable{\arabic{table}}

\begin{abstract}
We show that the tracking system in a collider detector can be used to efficiently identify boosted massive particles from their QCD backgrounds. We examine variables defined with tracking information which are sensitive to jet radiation patterns, including charged particle multiplicity and N-subjettiness. These variables are barely correlated with variables sensitive to the hard splitting scale in the jet, such as the filtered jet mass. Therefore these two kinds of variables should be combined to optimize the discriminating power. We illustrate the method with $W$ jet tagging. It is shown that for jet $p_T=500\gev$, one can gain a factor of 1.6 in statistical significance by combining filtered jet mass and charged particle multiplicity, over filtered mass alone. Adding N-subjettiness increases the factor to 1.8.
\end{abstract}

\maketitle
\thispagestyle{empty}

\pagenumbering{arabic}
\section{Introduction}

\label{sec:introduction}
Highly boosted massive standard model (SM) particles are important probes to TeV scale new physics at the Large Hadron Collider (LHC). These particles include the $W$ and $Z$ gauge bosons, the top quark, and possibly the Higgs boson if it is light. For example, it is essential to measure how the $WW$ scattering cross section grows with increasing center of mass energy, in which case boosted $W$ bosons are involved \cite{Chanowitz:1985hj, Butterworth:2002tt}. When these particles decay hadronically, we have to identify them from their QCD backgrounds. See Refs.~\cite{filtering, planarflow, pruning, Hackstein:2010wk, Katz:2010mr, Plehn:2009rk, Kribs:2009yh, Soper:2010xk, Chen:2010wk, Falkowski:2010hi, Kribs:2010hp, Katz:2010iq, Thaler:2008ju, Kaplan:2008ie, Almeida:2008tp, Krohn:2009wm, pull, Plehn:2010st, Bhattacherjee:2010za, Rehermann:2010vq, restframe-nsubjettiness, nsubjettiness, wtag, atlas, cms, Hook:2011cq, Plehn:2011sj} for previous studies. For this purpose, a large boost brings us both advantages and disadvantages. On the one hand, since the decay products are collimated, they are often clustered into a single jet and we are exempted from the combinatorial problem associated with unboosted particles. Sometimes it is even convenient to use a large jet radius to group as many as possible such particles to single fat jets \cite{filtering}. We will call these jets $W$/$Z$/top jets and in general, {\it boosted massive particle jets} or simply {\it boosted jets}. On the other hand,  since they behave as a single jet, we need to distinguish them from high $p_T$ QCD jets, namely, jets initiated from a high $p_T$ quark or gluon. Because of QCD radiation, the jet mass alone is not a good discriminant especially when we choose a large jet radius. 

To distinguish massive particle jets from QCD jets, we can utilize two differences between them. First, compared with QCD splittings, which are usually hierarchical, the momentum of a boosted massive particle is more evenly distributed among its decay products. This results in more than one hard subjet within the fat jet for a massive particle, while only one hard subjet for most QCD jets. Several algorithms \cite{filtering, pruning, trimming} have been invented to identify subjets and ``groom'' the fat jet by discarding the soft subjets. After grooming, the mass of a boosted jet remains close to the decaying particle's mass, while that of a QCD jet usually becomes very small. This allows us to use a mass window cut to eliminate most QCD jets and retain the massive particles we are interested in. To get the characteristic mass of a particular particle, it is important we measure the momenta of all stable particles, which is possible only with calorimeters. Therefore, in most previous jet substructure studies, hadronic calorimeter (HCAL) granularity is assumed. 

The second difference between boosted jets and QCD jets stems from their different color structures: QCD jets are initiated from a colored particle, while $W$/$Z$/Higgs bosons are color singlets. The color flow of the top decay is also different from a generic QCD jet with three hard subjets. This difference results in different radiation patterns, which are manifest in jet shape variables such as planar flow \cite{planarflow}, pull \cite{pull}, N-subjettiness \cite{nsubjettiness} and $R$-cores \cite{wtag}. Unlike the jet mass, to define and examine these variables, we do not need to have the information of all stable particles. In particular, we can take advantage of the tracking system, which has much finer granularity as well as better momentum resolution for low to moderately high momenta. Obviously, the disadvantage of using the tracking information is it is only available for charged particles whose fraction fluctuates. Therefore, it is complementary to calorimeter information. 

In this article, taking $W$ jets as an example, we discuss how to use tracking information to distinguish boosted massive particle jets from QCD jets. We explore variables sensitive to jets' radiation patterns. The simplest such variable is charged particle multiplicity, which nonetheless shows excellent discriminating power. All previously defined jet shape variables can also be calculated with charged particles alone. In particular, we will see N-subjettiness is very useful. These variables are not very sensitive to the mass scale of the particle decay or QCD splitting, and they should be combined with a jet grooming algorithm to optimize the discriminating power. Therefore, we examine these variables for jets containing two hard subjets (identified with the filtering algorithm) with masses close to the $W$ mass. The performances of these variables are conveniently quantified by the significance improvement characteristic (SIC) which is defined as \cite{sic}
\begin{equation}
\text{SIC} \equiv\varepsilon_S/\sqrt{\varepsilon_B},
\end{equation}
where $\varepsilon_S$ and $\varepsilon_B$ are respectively the signal ($W$ jets) and background (QCD jets) efficiencies.
It is shown that for the LHC, by combining charged multiplicity (or N-subjettiness) with filtering, we achieve an SIC of $\sim 1.6$ over the filtering method alone. This approach also gives better discriminating power than methods that combine correlated variables, such as in Ref.~\cite{Soper:2010xk}, where different jet grooming algorithms are combined. Combining filtered mass, charged multiplicity and N-subjettiness all together, we can reach an SIC of 1.8.  

The article is organized as follows. In Sec.~\ref{sec:ee}, we illustrate the difference between $W$ jets and QCD jets at $e^+e^-$ machines, which provide a clean environment without contaminations from initial state radiation and the underlying event. In Sec.~\ref{sec:lhc}, we examine jet substructure variables at the LHC and quantify their performances. We discuss our results in Sec.~\ref{sec:discussion} and conclude in Sec.~\ref{sec:conclusion}.

\section{W jets and QCD jets at $e^+e^-$ machines}
\label{sec:ee}
We first compare W jets with QCD jets produced at $e^+e^-$ machines, where jet substructure is not contaminated by initial state radiation and the underlying event. The lessons learned in this section can be easily adapted for the LHC, which we discuss in the next section.
\subsection{Charged particle multiplicity}
Charged particle multiplicity is a simple but powerful variable that is sensitive to the jet color structure. For example, it can be used to distinguish gluon jets from quark jets \cite{quark-gluon}. Average charged particle multiplicities ($\langle\nch\rangle$) in inclusive hadronic events have been measured at a number of $e^+e^-$ machines. The results are compiled in Ref.~\cite{ee_exp} and reproduced in Fig.~\ref{fig:ee_exp}. These measurements include a variety of center of mass energies from $12\gev$ at PETRA, up to LEP 2 energies. Theoretically, the absolute value of $\langle\nch\rangle$ cannot be predicted because it involves non-perturbative physics. However, its scaling can be described by the modified leading log approximation (MLLA) and local parton-hadron duality (LPHD) \cite{mlla}, see Ref.~\cite{nch_review} for a review. 
The prediction from MLLA+LPHD is shown in Fig.~\ref{fig:ee_exp}, together with the prediction from Pythia 8 \cite{pythia8} simulations. We see that the MLLA+LPHD fit and the Pythia 8 prediction are almost identical, which agree with data excellently. 

In Fig.~\ref{fig:ee_exp}, we also notice that charged particle multiplicity grows slowly especially at high energies. This is an example of the fact that the radiation pattern is more sensitive to the color structure than the energy scale. This effect has important consequence as will be seen in our study of jet substructure.   
\begin{figure}[th!]
\begin{center}
\includegraphics[width=0.7\textwidth]{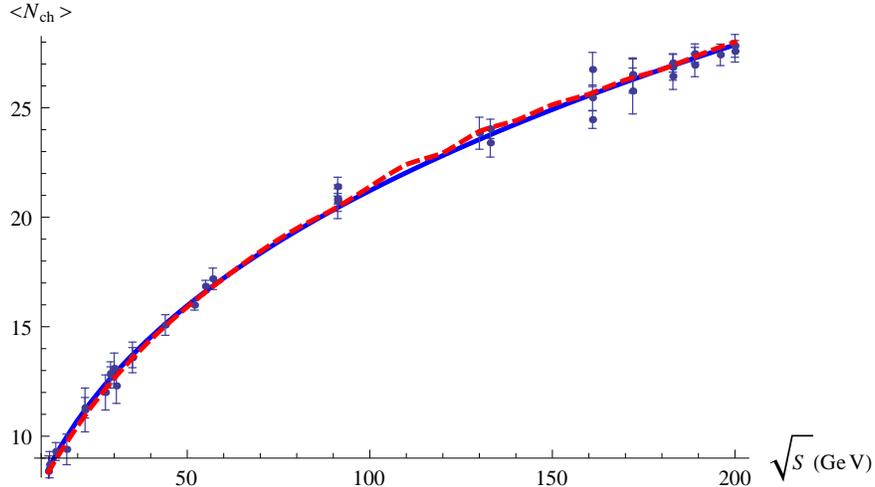} 
\caption{Average charged particle multiplicities as a function of center of mass energy measured at $e^+e^-$ machines (points) \cite{ee_exp}, together with the MLLA+LPHD fit (blue solid) and Pythia 8 simulations (red dashed).}
\label{fig:ee_exp}
\end{center}
\end{figure}

For the $W$ boson, a color singlet particle, the average charged multiplicity should be the same as in Fig.~\ref{fig:ee_exp} at $\sqrt{s}=M_W$, which was confirmed at LEP 2 \cite{nch_w}. Not shown in Fig.~\ref{fig:ee_exp} is the dispersion of the charged multiplicity (defined as $\langle(\nch-\langle\nch\rangle)^2\rangle^{1/2}$), which is $\sim6$ for the $W$ boson. Except for experimental effects, the charged particle multiplicity distribution is invariant under the boost to the $W$ boson, which allows us to use it to identify $W$ jets.

The inclusive charged multiplicity is not directly applicable for identifying $W$ jets since jets are defined for a finite spatial region. We then examine charged multiplicities for individual jets using Pythia 8 simulations. As mentioned in the introduction, a QCD jet without a hard splitting can be easily distinguished from a $W$ jet by using the jet grooming algorithms, therefore, we focus on QCD jets with a hard splitting (2-prong QCD jets), which mimic a $W$ jet more closely. Nonetheless, we will also include 1-prong QCD jets for comparison. For illustration, we consider $e^+e^-\rightarrow W^+W^-\rightarrow q\bar ql\nu$ events and $e^+e^-\rightarrow q\bar q g$ events in the following fixed configurations (Fig.~\ref{fig:special_kinematics}): the hadronically decaying $W$ moves along the $x$ axis which decays to two quarks with symmetric 4-momenta,
\begin{equation}
p_q=\left(\frac{\sqrt{p_T^2+M_W^2}}{2}, \frac{p_T}{2}, 0, \frac{M_W}{2}\right), \  \ p_{\bar q}=\left(\frac{\sqrt{p_T^2+M_W^2}}{2}, \frac{p_T}{2}, 0, -\frac{M_W}{2}\right), \label{eq:qq}
\end{equation}
where $p_T$ is fixed to be $500\gev$.
For $e^+e^-\rightarrow q\bar q g$, we choose the quark and the  gluon to mimic the partons from a $W$ decay, therefore, the 4-momenta are
 \begin{eqnarray}
&p_q=\left(\frac{\sqrt{p_T^2+M_W^2}}{2}, \frac{p_T}{2}, 0, \frac{M_W}{2}\right), \  \ p_g=\left(\frac{\sqrt{p_T^2+M_W^2}}{2}, \frac{p_T}{2}, 0, -\frac{M_W}{2}\right),\nonumber\\
 & p_{\bar q}= (p_T, -p_T, 0, 0).
\end{eqnarray}

\begin{figure}[th!]
\begin{center}
\includegraphics[width=0.7\textwidth]{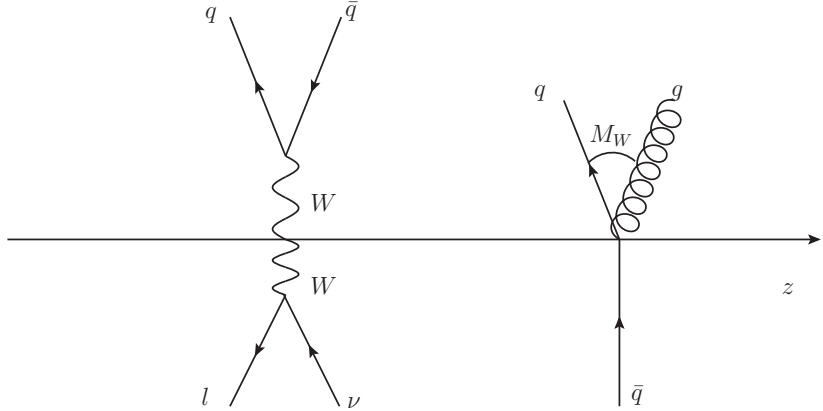} 
\caption{Fixed momentum configurations for illustration. Left: $WW\rightarrow q\bar ql\nu$ with the hadronic $W$ moving perpendicularly to the beam and decaying to quarks with symmetric momenta. Right:   $e^+e^-\rightarrow q\bar q g$ with a $qg$ pair mimicking a $W$ boson. }
\label{fig:special_kinematics}
\end{center}
\end{figure}

There is another configuration with the above momenta in which the two quarks are close to each other and the gluon is in the opposite hemisphere. This configuration happens much less often than the one showing in Fig.~\ref{fig:special_kinematics}, and we have ignored it in our illustration. In a realistic situation at the LHC as will be discussed in Sec.~\ref{sec:lhc}, we should include all possible configurations contributing to a high $p_T$ jet. This is more conveniently done with Pythia 8 or other simulations. 

We keep the momentum and color flow configurations fixed as in Fig.~\ref{fig:special_kinematics}, and repeatedly use Pythia 8 to simulate showering and hadronization and obtain two data samples corresponding to the two processes. We cluster stable particles to anti-$k_t$ jets ($R$=1.2) with FastJet \cite{fastjet}. Each $WW$ event then contains a $W$ jet in the upper hemisphere, and each $q\bar qg$ event contains a 2-prong jet in the upper hemisphere and a 1-prong jet in the lower hemisphere. No cut is used in this procedure. The average number of charged particles for the $W$ jet, the 2-prong and 1-prong QCD jets are shown in Fig.~\ref{fig:nch_ee}. From Fig.~\ref{fig:nch_ee}, we see the average charged multiplicity of a $W$ jet is larger than that of a 1-prong QCD jet while smaller than a 2-prong QCD jet. This is due to their different color structure. In particular, the 2-prong QCD jet is color connected to the other side of the event, therefore it contains more radiation than the $W$ jet. To distinguish a $W$ jet from a 2-prong QCD jet, we can apply a cut $\nch\leq \nch^{\cut}$. For example, when $\nch^{\cut}= 19$, we keep 63\% $W$ jets and 7.7\% 2-prong QCD jets, which boosts the SIC by a factor of 2.3. Because of the large boost and the large jet radius, $R=1.2$, almost all particles from the $W$ decay are included in the $W$ jet.  Due to charge conservation, the $W$ jet (almost) always contains an odd number of charged particles. If we keep only jets with odd number of charged particles, we obtain a larger SIC. However, this feature is easily lost due to experimental acceptance, and we do not pursue this possibility further.  
\begin{figure}[th!]
\begin{center}
\includegraphics[width=0.7\textwidth]{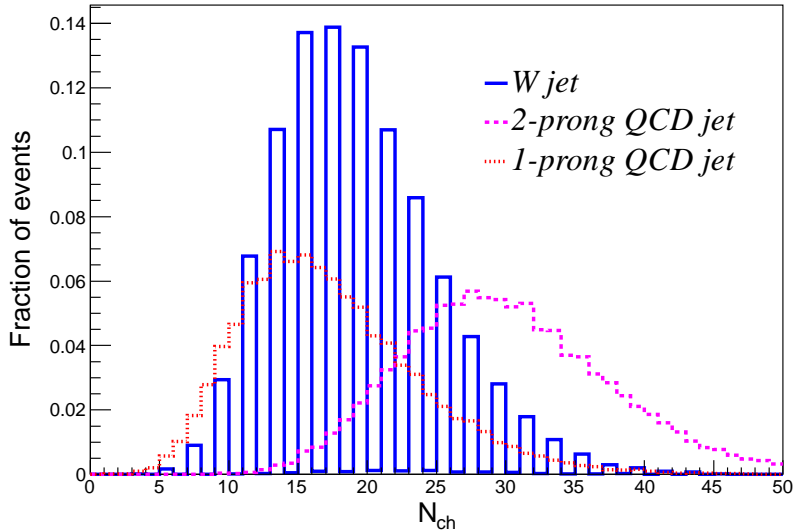} 
\caption{Charged particle multiplicities for $W$ jets, 2-prong and 1-prong QCD jets with $p_T=500\gev$, in the fixed momentum configurations of Fig.~\ref{fig:special_kinematics} (see text).}
\label{fig:nch_ee}
\end{center}
\end{figure}

\subsection{N-subjettiness} 
Other existing jet shape variables can be defined with charged particles too. Here, we take N-subjettiness as an example, which is derived from N-jettiness \cite{njettiness} and defined in Ref.~\cite{nsubjettiness} as follows. 
For a set of particles $\{ i\}$ in a jet with radius $R_0$ and a set of $N$ axes $\{J\}$, we define the distance $\Delta R_{J,i}\equiv \sqrt{\Delta\eta_{J,i}^2+\Delta\phi_{J,i}^2}$ for each $(J,i)$ pair. Then we define a quantity
\begin{equation}
\tilde{\tau}_N^{(\beta)}\equiv\frac{1}{d_0}\sum_ip_{T,i}\min\{\Delta R_{1,i}^\beta, \ldots \Delta R_{J,i}^\beta \ldots \Delta R_{N,i}^\beta\}, \label{eq:tautilde}
\end{equation}
where $d_0=\sum_ip_{T,i}R_0^\beta$ and $\beta$ is a pre-selected constant. We vary the directions of the $N$ axes to find the minimum $\tilde\tau_N^{(\beta)}$, which is defined as $N$-subjettiness, $\tau_N^{(\beta) }$. 

In our example with fixed momentum configuration, we can simply use the momenta in Eqs.~(\ref{eq:qq}) as the two axes to calculate $\tilde\tau_2^{(\beta)}$, which does not differ significantly from the true $\tau_2^{(\beta)}$ after minimization. The $\tau_2^{(\beta)}$ distributions for $W$ jets and 2-prong QCD jets are shown in Fig.~\ref{fig:nsub_ee}, where we have chosen $\beta = 1$. The variable $\tau_2$ quantifies how likely a given jet contains two hard subjets, a smaller value corresponding to a larger likelihood. In our special momentum configuration, we always have two hard subjets. In this case, $\tau_2$ becomes a measure of how diffuse the radiation is. As expected, more often $\tau_2$ is larger for 2-prong QCD jets than $W$ jets, which allows us to apply a cut, $\tau_2 < \tau_2^{\text{cut}}$, to suppress background QCD jets.
\begin{figure}[th!]
\begin{center}
\includegraphics[width=0.7\textwidth]{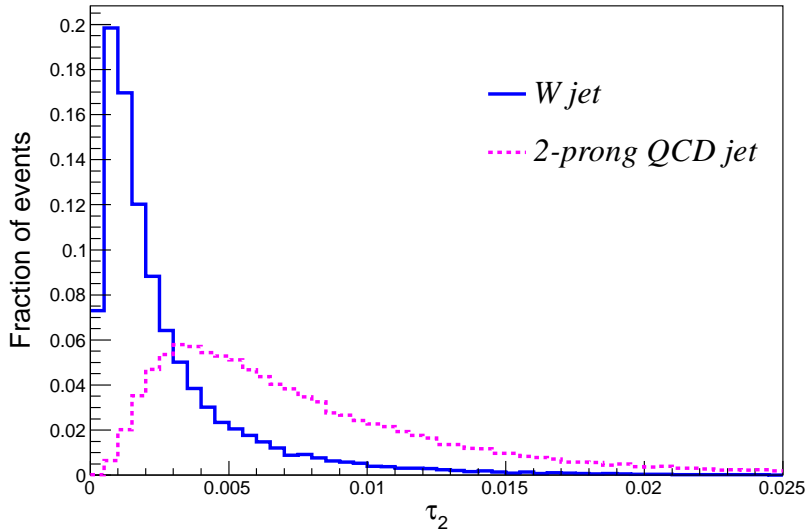} 
\caption{2-subjettiness with axes fixed to Eq.~(\ref{eq:qq}), for jets with $p_T=500\gev$ in the fixed momentum configurations of Fig.~\ref{fig:special_kinematics}.}
\label{fig:nsub_ee}
\end{center}
\end{figure}
 In the general case, when the jet momentum configuration is not fixed, a better variable is $\tau_2/\tau_1$ \cite{nsubjettiness}, which will be discussed in the next section for the LHC.
 
\section{$W$ jet tagging at the LHC}
\label{sec:lhc}
We now turn to the LHC, where $W$ jet tagging is much more challenging due to presence of initial state radiation and the underlying event. As an example, we consider $W$ jets and QCD jets in the $p_T$ range $(500, 550)\gev$. These jets are obtained as in Ref.~\cite{wtag}: we use Pythia 8 to generate high $p_T$ $WW$ pairs which decay semiletonically, and $W$+jet events with the $W$ decaying leptonically.  Each $WW$ event then contains a $W$ jet, while each $W$+jet event contains a high $p_T$ QCD jet. The visible stable particles in these events are grouped in $0.1\times 0.1$ bins in the $(\eta, \phi)$ plane, corresponding to the HCAL granularity. Jets are found with the Cambridge/Aachen algorithm ($R=1.2$). Since we are interested in jets with a hard splitting and mass close to the $W$ mass, we apply the filtering algorithm with the mass drop method and examine further jets passing the filtered jet mass ($\mfilt$) cut, $(60, 100)\gev$. The mass drop parameters are $\mu=0.71$ and $y_\cut =0.09$, which give a factor of 2.2 improvement in the SIC over the original fat jets. With this cut, we reduce the number of QCD jets by 91\% and at the same time retain 66\% $W$ jets. Events passing this cut all have a hard splitting, therefore we can apply the lessons learned from the previous section. 

We identify tracks with $p_T>1\gev$ and $|\eta|<2.5$ in a given jet, and use them to calculate the charged particle multiplicity and N-subjettiness. Note that although the jet has passed the filtered mass window cut, we include all tracks in the original fat jet, which give us more information about the jet color structure. The most efficient N-subjettiness variable for $W$-tagging is $\tau_2/\tau_1$, where we have set $\beta = 1$ in Eq.~(\ref{eq:tautilde}) and used the code in Ref.~\cite{nsub_code}. In addition, the filtered jet mass distributions still differ for $W$ jets and QCD jets after imposing the $(60, 100)\gev$ mass window cuts. We therefore include the filtered mass as well. The three variables under consideration are shown in Fig.~\ref{fig:vars_lhc}. From Fig.~\ref{fig:vars_lhc}, we see $\nch$ and $\tau_2/\tau_1$ have similar features as in a $e^+e^-$ machine (Fig.~\ref{fig:nch_ee} and Fig.~\ref{fig:nsub_ee}), but the distinctions between QCD jets and $W$ jets are smaller.

\begin{figure}
\begin{center}
\begin{tabular}{cc}
\includegraphics[width=0.5\textwidth]{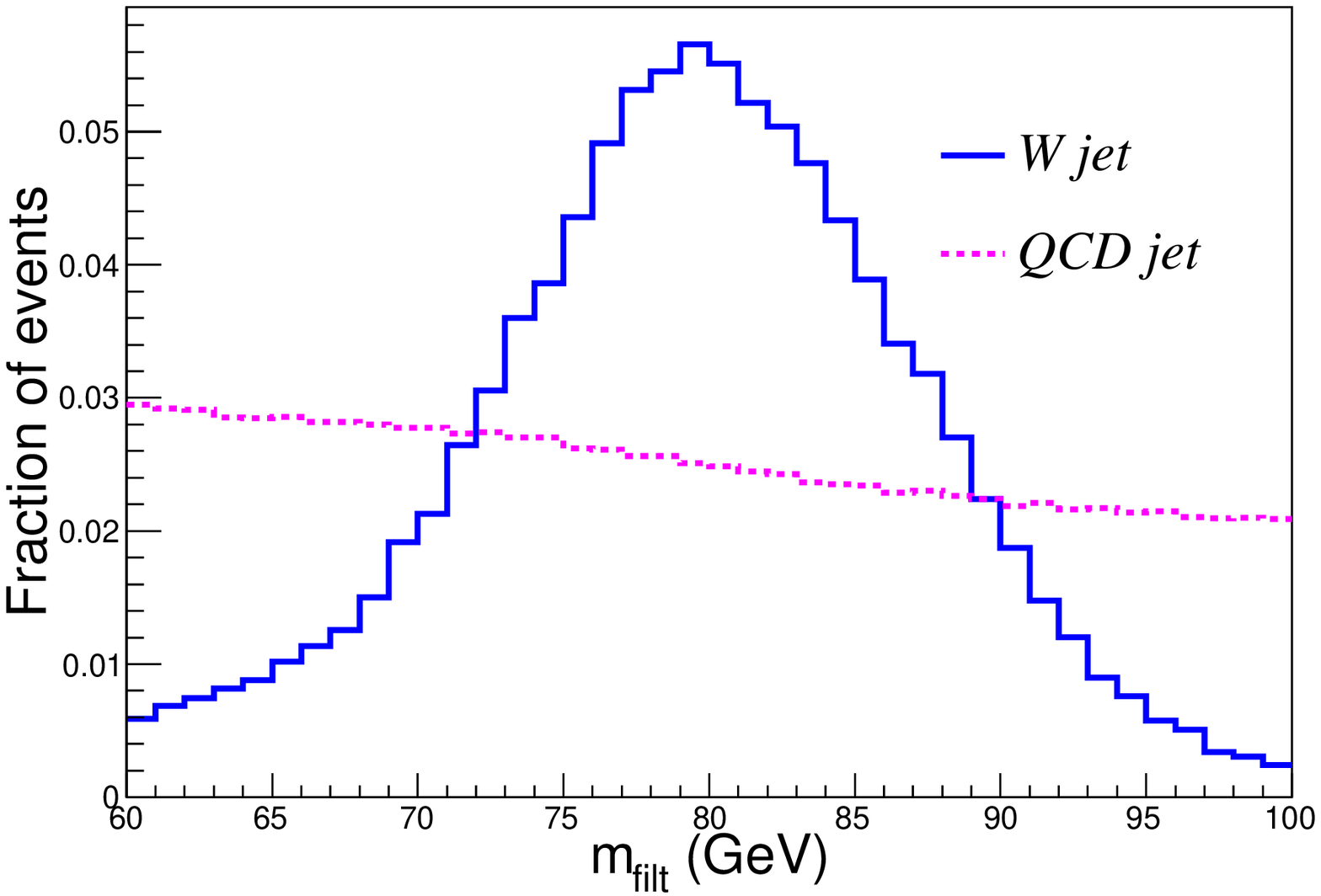}
&\includegraphics[width=0.5\textwidth]{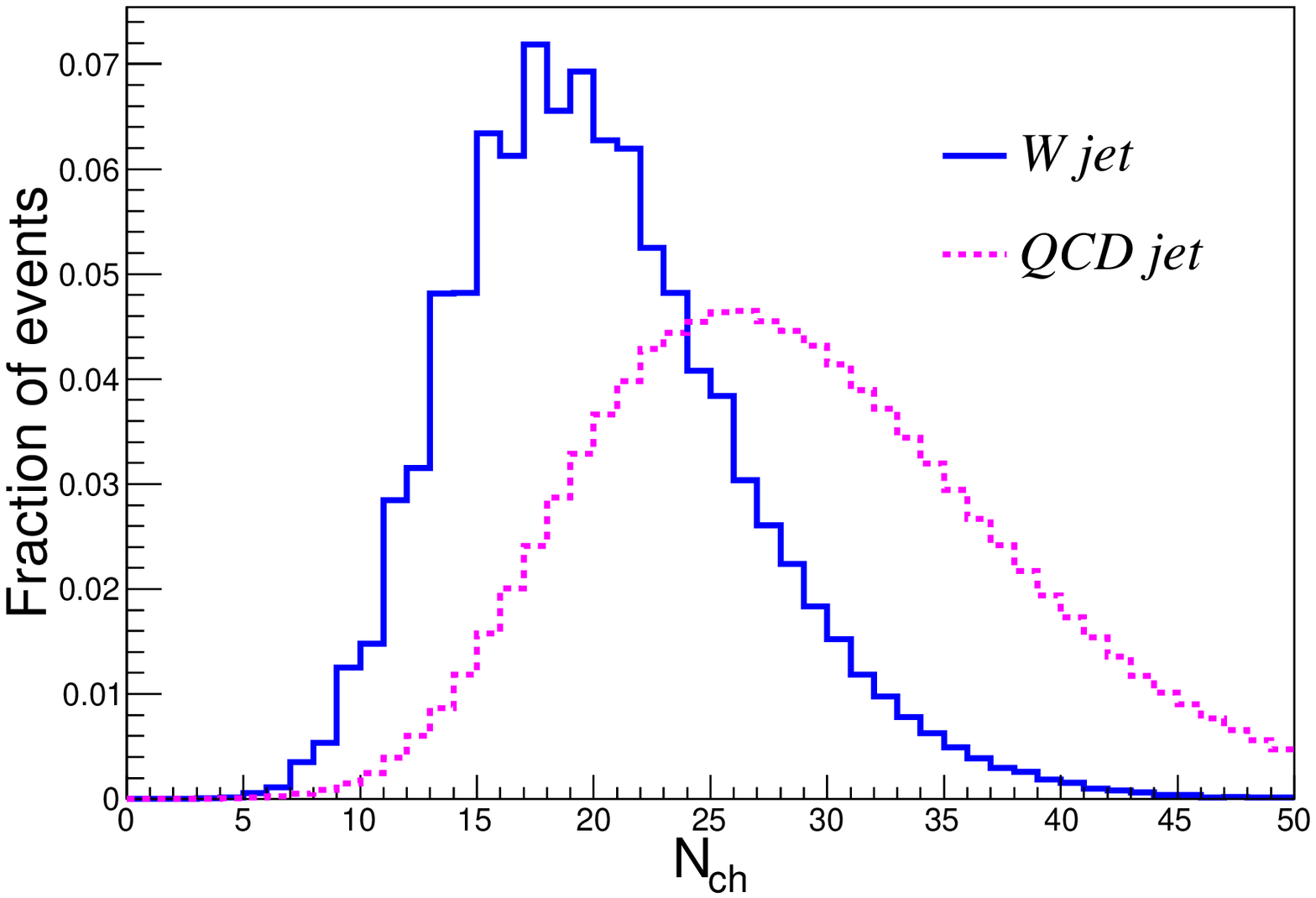}
\end{tabular}
\includegraphics[width=0.5\textwidth]{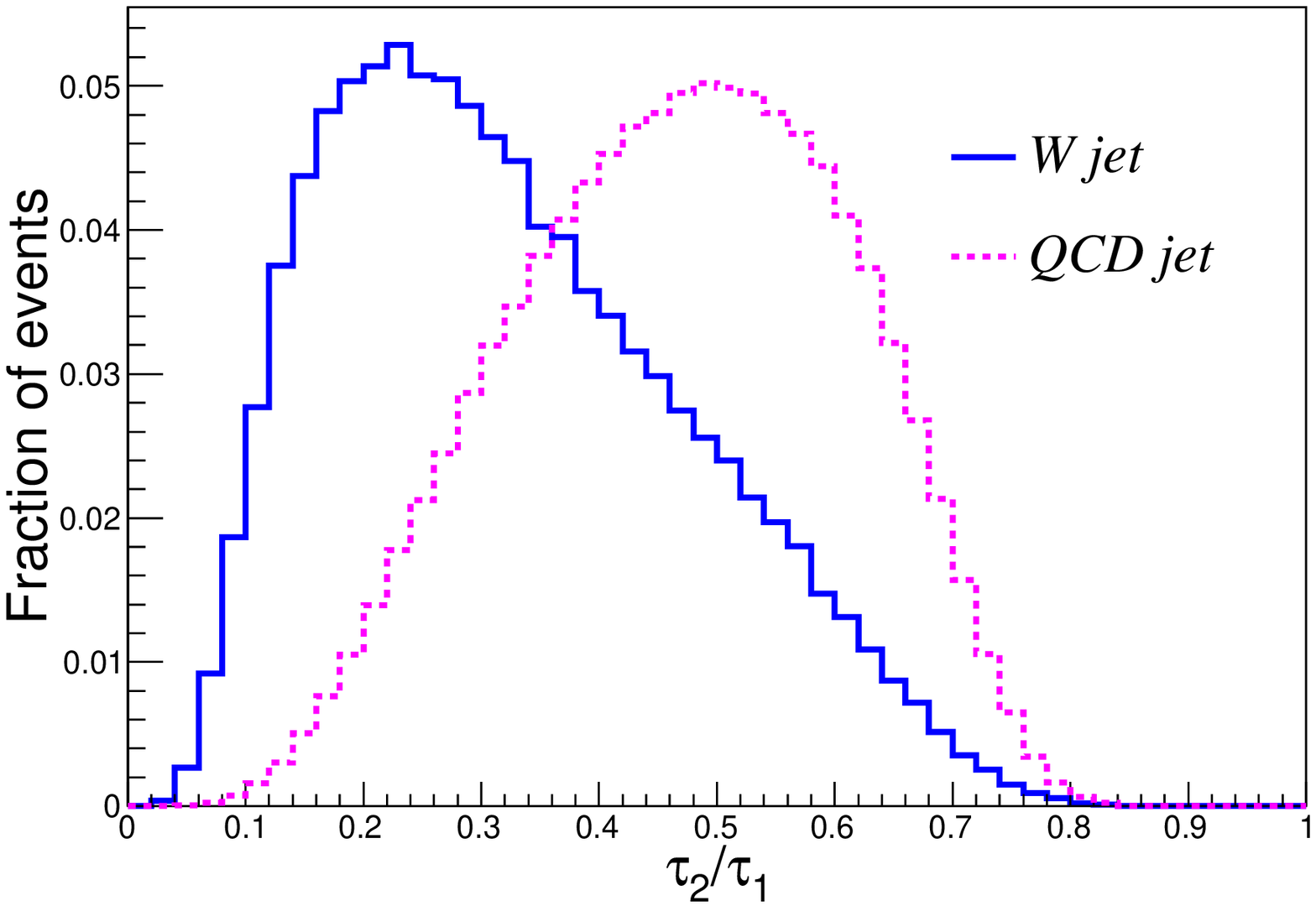}
\caption{Variables considered for the LHC: jet mass after filtering, charged multiplicity and $\tau_2/\tau_1$, for $p_T^\jet\in(500, 550)\gev$. The filtered mass is constructed assuming HCAL granularity; $\nch$ and $\tau_2/\tau_1$ are constructed using tracks with $p_T>1\gev$ and $|\eta|<2.5$.   \label{fig:vars_lhc}}
\end{center}
\end{figure}

For a single variable, we can apply a rectangular cut to improve the SIC. For two or more variables, it is better to combine them using a multivariate classifier such as the Boosted Decision Trees method \cite{bdt} in the package TMVA \cite{TMVA}. The best performances for the three individual variables as well as all combinations are given in Tab.~\ref{tab:significance}.

\begin{table}[t]
\begin{center}
\begin{tabular}{|c|c|c|c|}
\hline
& $\mfilt$ & $\  \ \nch \  \ $ &  $\  \  \tau_2/\tau_1 \  \  $ \\
\hline
$\mfilt$ & 1.15 & 1.66 (1.59) & 1.67 (1.58)\\
\hline
$\nch$     & - & 1.34 & 1.55 (1.50)\\
\hline
$\tau_2/\tau_1$& - & -   & 1.39\\
\hline
all:&\multicolumn{3}{|c|}{1.85}\\
\hline 
\end{tabular}
\end{center}
\caption{Optimized improvement in the SIC. The events have passed an overall filtered mass cut $(60, 100)\gev$. The diagonal elements of the first three rows are obtained by using individual variables with an optimized rectangular cut. The off-diagonal elements are obtained by combining a pair of variables: the numbers in the parentheses are obtained using rectangular cuts and the numbers outside are from BDT. The best improvement for combining all three variables in BDT is given in the last row. \label{tab:significance}}
\end{table}
We emphasize here the numbers in Tab.~\ref{tab:significance} are obtained on jet samples that have passed the filtered mass cut, $(60, 100) \gev$, which has already increased the SIC by a factor of 2.2. Therefore, the overall improvement is the number in Tab.~\ref{tab:significance} multiplied by 2.2. We see the number for using filtered mass alone is 1.15, which means $(60, 100) \gev$ is not the optimum mass window. The extra factor of 1.15 corresponds to a narrower mass window $(72, 92)\gev$ and the best improvement one can get from filtering alone is 2.5. Better improvements are obtained from $\nch$ or $\tau_2/\tau_1$ (with the filtered mass window fixed to $(60, 100)\gev$): we obtain 1.34 (1.39) by optimizing the cut on $\nch$ ($\tau_2/\tau_1$).

\begin{table}[t]
\begin{center}
\begin{tabular}{|c|c|c|c|}
\hline
&$m_\filt$&$\nch$&$\tau_2/\tau_1$\\
\hline
$m_\filt$&1&-0.08&-0.12\\
\hline
$\nch$&-0.08 & 1 &0.51 \\
\hline
$\tau_2/\tau_1$&-0.12 &0.51&1\\
\hline
\end{tabular}
\quad\quad
\begin{tabular}{|c|c|c|c|}
\hline
&$m_\filt$&$\nch$&$\tau_2/\tau_1$\\
\hline
$m_\filt$&1  &0.07&-0.14\\
\hline
$\nch$&0.07 & 1 &0.50 \\
\hline
$\tau_2/\tau_1$&-0.14 &0.50&1\\
\hline
\end{tabular}
\end{center}
\caption{Linear correlation matrices of the variables. Left: $W$ jets; right: QCD jets. \label{tab:correlations}}
\end{table}
  
One may also be interested in the performance of a single variable sensitive to the radiation, such as $\tau_2/\tau_1$, without imposing a filtered mass cut. It turns out one can improve the significance by a factor of 1.44 using $\tau_2/\tau_1$ calculated from charged particles \footnote{The performance depends on the value of $\beta$ in Eq.~(\ref{eq:tautilde}). It turns out that with a filtered mass cut, $\beta = 1$ is a better choice than $\beta = 2$. Therefore we have used $\beta = 1$ all through the paper. However, without a filtered mass cut, $\beta =2$ works better, which gives a larger SIC of 1.58 for jet $p_T=500\gev$. A detailed study of the $\beta$ dependence is beyond the scope of the article.} for jets with $R=1.2$ and $p_T=500\gev$. This is lower than what we get from filtering. Nevertheless, $\tau_2/\tau_1$ may be more useful if we do not know the exact mass of the boosted particle. Also at higher $p_T$'s, the resolution for mass measurement degrades, while we expect radiation variables to work better. This is because for higher $p_T$'s, the decay products of a color singlet particle occupy a smaller region while the radiation pattern of a QCD jet does not change significantly, which make the two cases more distinguishable. Similar observation has been made in Ref.~\cite{wtag}. In the extreme case, all decay products of a color singlet particle enters a single or a few adjacent calorimeter cells which makes the mass information unavailable, and we are forced into an inclusive search of color singlet particles without using their mass information.

 We can obtain better discriminating power if we combine two variables and vary the cuts on both of them. From Table \ref{tab:significance}, we see a factor of $\sim1.6$ is reached if we optimize rectangular cuts on both $\mfilt$ and $\nch$ ($\tau_2/\tau_1$), or if we combine them in BDT, the latter being slightly better. We also notice combining $\nch$ and $\tau_2/\tau_1$ gives us smaller improvement (1.55) than combining one of them with the filtered mass, despite the fact that each alone is an excellent discriminant. This is due to the larger correlation between $\nch$ and $\tau_2/\tau_1$, both of which measure the amount of radiation in the jets. On the other hand, the correlation between $\nch$ ($\tau_2/\tau_1$) and $m_\filt$ is small. These correlations are manifest in the two dimensional distributions for each pair of variables, shown in Fig.~\ref{fig:2d}. The linear correlation matrices of the three variables are given in Tab.~\ref{tab:correlations}. 

From the above results, we draw the conclusion that if we are to use two variables for $W$ jet tagging, the best way is to choose one variable sensitive to the hard splitting scale and the other one sensitive to the amount of radiation. These two kinds of variables characterize two major differences between $W$ jets and QCD jets and they are barely correlated. This is in accordance with the observation we made in Sec.~\ref{sec:ee}, where we saw that charged multiplicity increase slowly with respect to the center of mass energy. By doing so, we obtain significant improvement than using each individual variable.

Finally, we can combine all three variables in BDT. Despite the sizable correlation between $\nch$ and $\tau_2/\tau_1$, they still contain different information which can improve the SIC when combined. We show the SIC as a function of signal efficiency in Fig.~\ref{fig:3vars}. From Fig.~\ref{fig:3vars}, we see that a factor of 1.85 is achieved using the optimum cut, with sizable signal efficiency of $\sim 0.3$. We may also add other variables as in Ref.~\cite{wtag}, and the SIC will gradually saturate. In Ref.~\cite{wtag}, a set of 25 variables are used which yield a factor of 2.4. It turns out by adding the two extra variables, $\nch, \tau_2/\tau_1$ to the set ($m_\filt$ was included in Ref.~\cite{wtag}), we only obtain a few percent improvement, which means most of the information is redundant. However, we emphasize that if one would like to sacrifice performance for simplicity and choose to use only a few variables, the variables we have considered in this article are among the best ones.

\begin{figure}[th!]
\begin{center}
\includegraphics[width=0.8\textwidth]{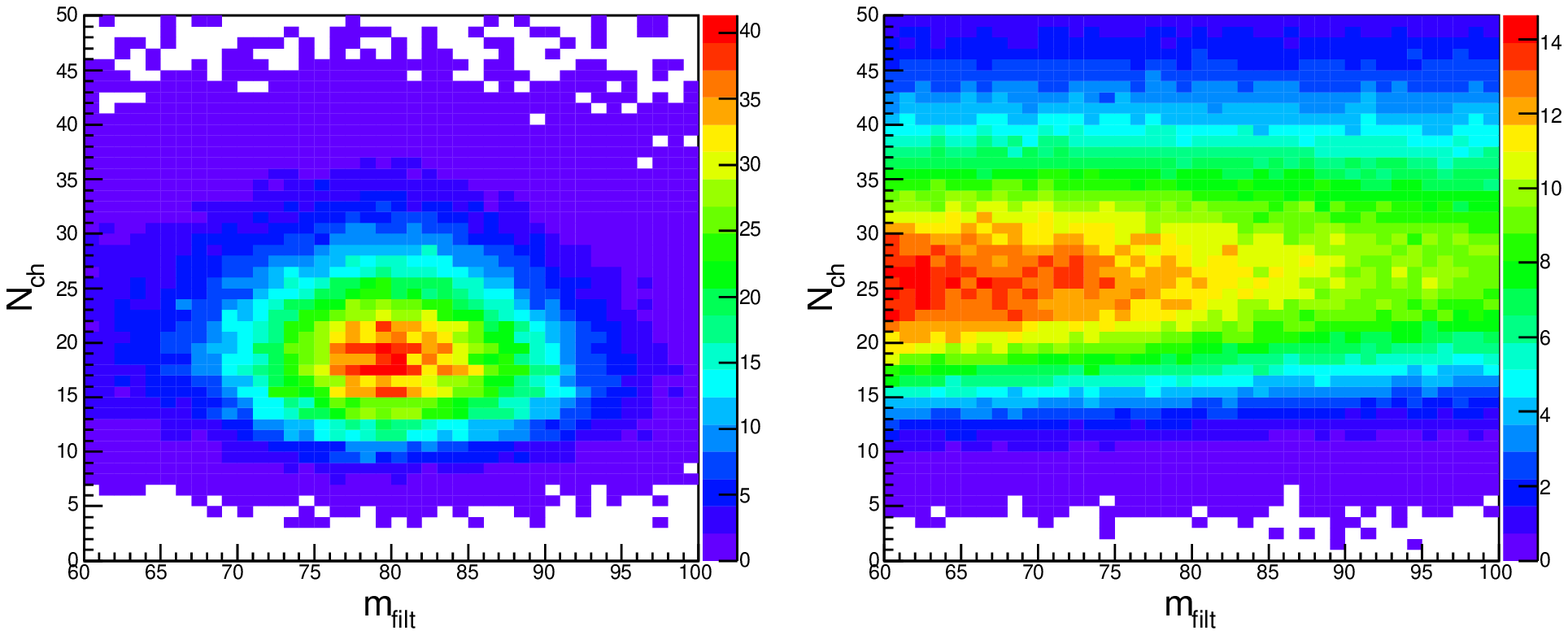} 
\includegraphics[width=0.8\textwidth]{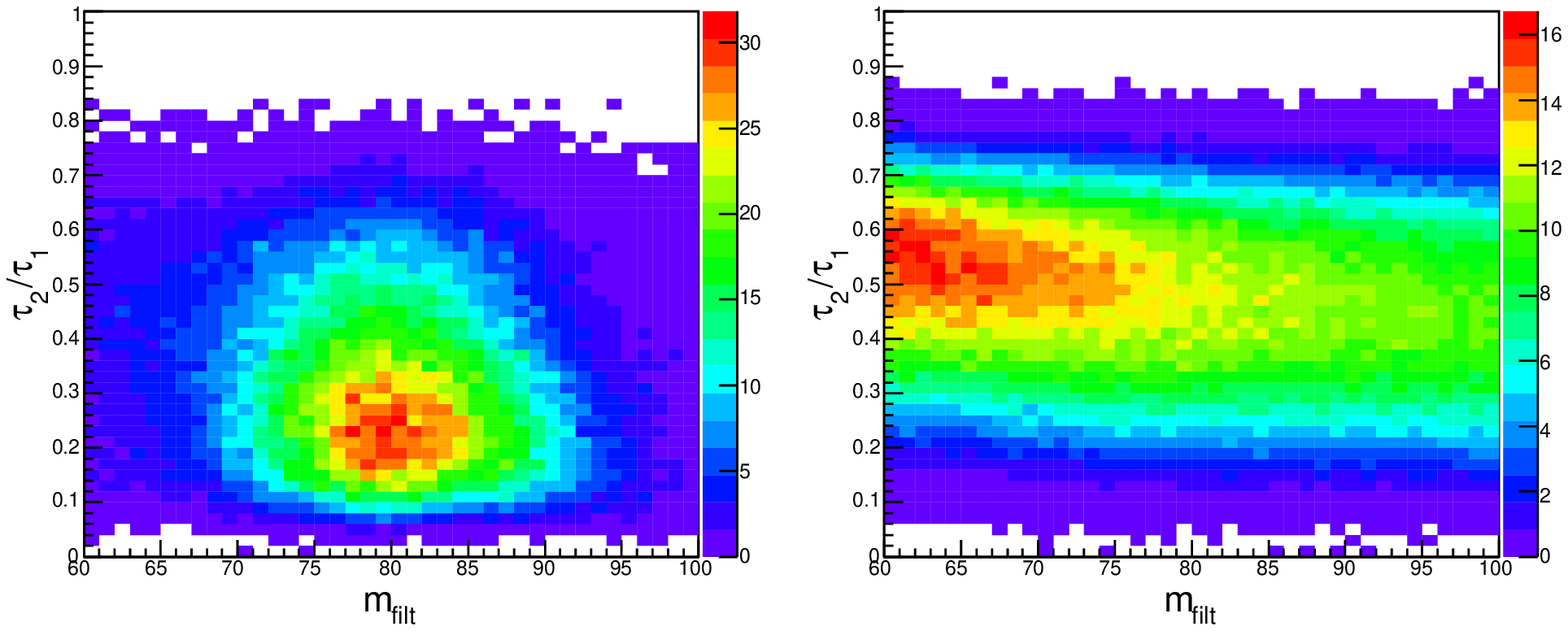} 
\includegraphics[width=0.8\textwidth]{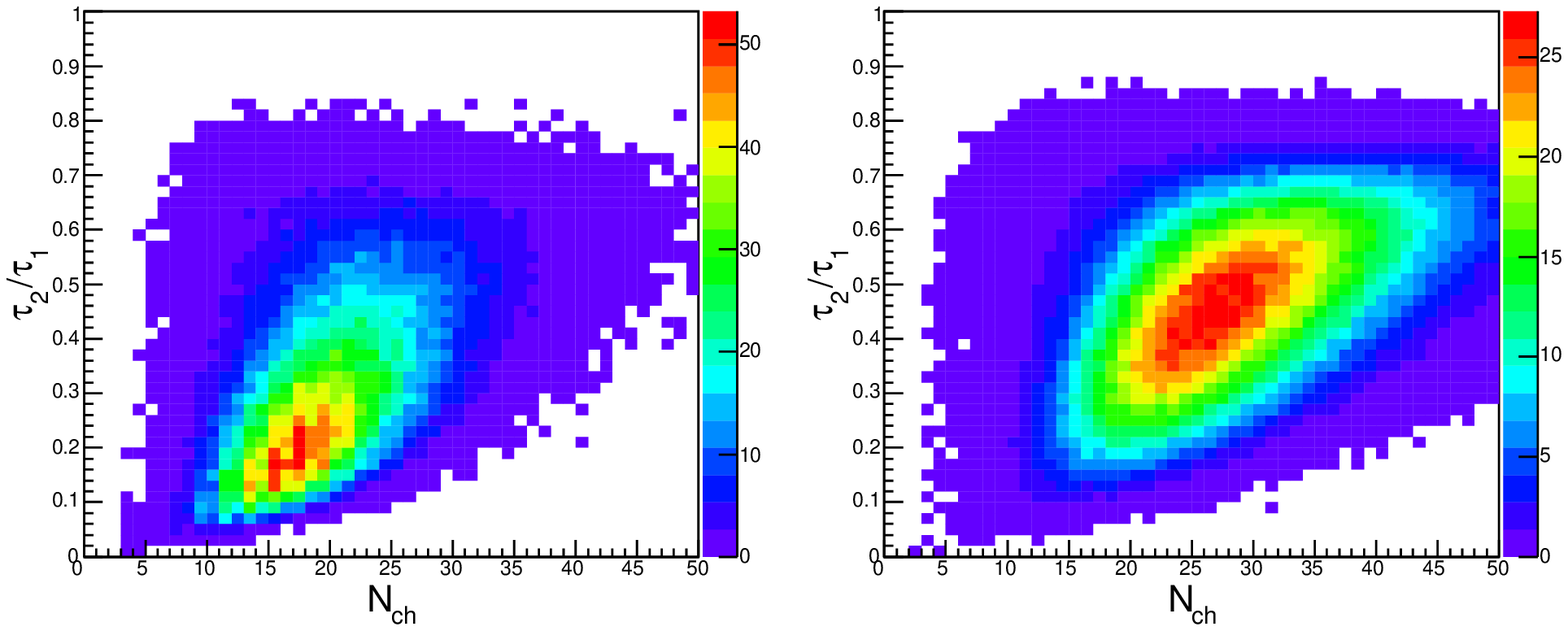} 
\caption{Two dimensional distributions for variable pairs. Left: $W$ jets; right: QCD jets. The number of events is normalized to 10k for each plot.\label{fig:2d}}
\end{center}
\end{figure}

\begin{figure}[th!]
\begin{center}
\includegraphics[width=0.6\textwidth]{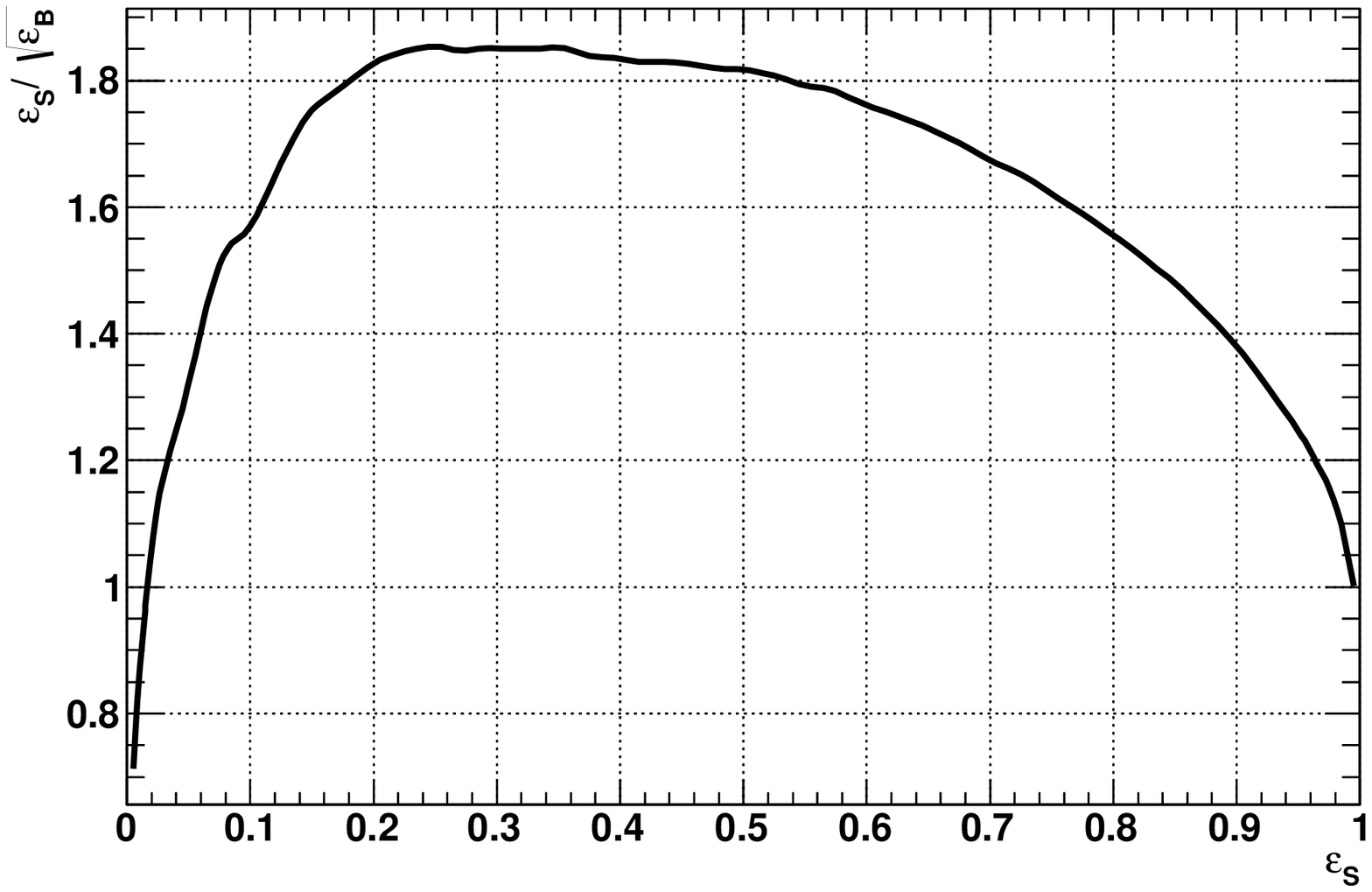} 
\caption{\label{fig:3vars} The SIC as a function of signal efficiency by combining $\mfilt$, $\nch$ and $\tau_2/\tau_1$ in BDT, for $p_T^\jet\in(500, 550)\gev$.}
\end{center}
\end{figure}  
\section{Discussions}
\label{sec:discussion}
\subsection{Experimental considerations}
In the above discussions, we have not taken into account the experimental efficiency for reconstructing tracks in a jet. In Ref.~\cite{atlas_tdr}, jets with $p_T\sim 200\gev$ are studied where it is shown the efficiency for identifying tracks is around 90\% with a fake rate about 0.1\%. The performance of our tracking variables will degrade accordingly. One may be concerned about whether the efficiency decrease significantly for higher $p_T$'s. This deserves dedicated studies using both simulations and the real data. However, we believe this is not the case for QCD jets. The reason is, the scaling of $\nch$ at a hadron collider is similar to that of $e^+e^-$ machines \cite{nch_pp} we see in Fig.~\ref{fig:ee_exp}. As we have noticed, the number of tracks grow very slowly at high energies. The angular distributions of these tracks will not change significantly either. Therefore, the efficiency will not change significantly for QCD jets. The $W$ jet is a different case: when the boost is larger, all the tracks will be packed in a smaller region, which may cause the efficiency to drop significantly. However, as we have used the fact $\nch$ is smaller for $W$ jets, a drop in efficiency will only make the number even smaller and will not hurt the discriminating power. An observation of dense tracks (or hits if tracks are difficult to reconstruct) in a small region combined with few tracks outside of that region demonstrates a very clean signal of a $W$ jet. 

Another concern about using high momentum tracks is the tracking resolution degrades for higher $p_T$'s.  This does not affect the charged multiplicity measurement, but affects N-subjettiness. However, since the momentum of the jet is shared by tens of particles, each charged particle does not usually have a very high $p_T$. For the $500\gev$ jets we considered in Sec.~\ref{sec:ee}, the leading track's $p_T$ is shown in Fig.~\ref{fig:pt}, from which we see the track's $p_T$ rarely goes above 200 GeV where the resolution of the momentum measurement is still better than 10\% \cite{atlas_tdr, cms_tdr}. Even in the presence of very high $p_T$ tracks, these tracks will likely dominate the directions of the subjets, and according to Eq.~(\ref{eq:tautilde}), not significantly contribute to N-subjettiness. Therefore, in this article we have ignored the experimental resolution which should not affect our results significantly unless we are interested in jets with extremely high $p_T$'s ($> 1$ TeV).

\begin{figure}[th!]
\begin{center}
\includegraphics[width=0.6\textwidth]{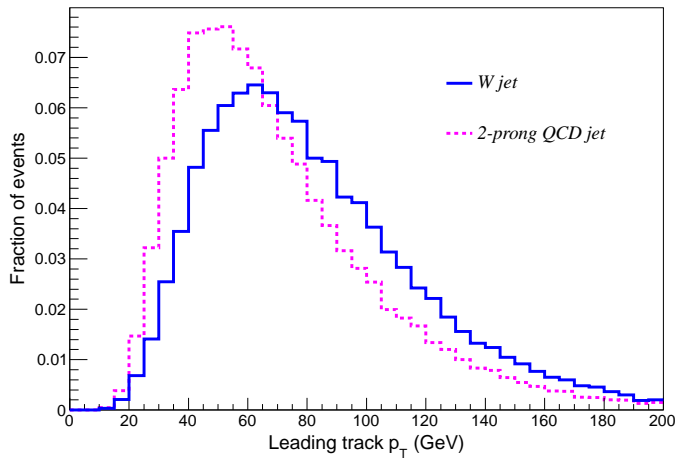} 
\caption{\label{fig:pt} The leading track's $p_T$ in jets with $p_T^\jet=500\gev$.}
\end{center}
\end{figure}  

Particle flow \cite{particle_flow} is an interesting and useful experimental approach, in which one combines the information from all subdetectors to reconstruct both charged and neutral particles. Very briefly, charged particles, including muons, electrons and charged hadrons, are first reconstructed and their energy removed from HCAL and ECAL clusters. The remaining clusters are identified as neutral particles including photons and neutral hadrons. Therefore, particles reconstructed from the particle flow algorithm (which we call PF particles) are a superset of the charged particles. If PF particles are used in jet reconstruction, the variables discussed in this paper can be naturally defined without explicitly referring to whether tracking or calorimeter information is used. For example, we can count the number of PF particles in stead of charged particles in our particle multiplicity definition. Since the PF particles are a superset of the charged particles, one may obtain better results than those quoted in this articles. Of course, for jets with a higher $p_T$, more particles are merged in single calorimeter clusters and it remains to be studied whether significant improvement can be achieved. 

\subsection{Other particles}
The same method can be used on other color singlet particles such as the $Z$ boson and a light Higgs boson. The major difference between these particles and the $W$ boson is in their masses.  As we have discussed, the small mass difference between $Z$ and $W$ does not affect significantly the charged multiplicity or N-subjettiness. Therefore, if we are only interested in one of them and treat the other one as a background, mass variables such as the filtered mass is necessary. However, as we have discussed in the previous section, for extremely high $p_T$ jets, one cannot measure accurately the mass of the color singlet particle, although it is easy to distinguish a color singlet particle from a QCD jet.

Color singlet particles also differ in their spins, which affect the angular distributions of the decay products with respect to the moving direction of the decaying particle. For a transverse vector boson, one of the fermions from the decay tends to go along the direction of the vector boson, while the other one goes against it in the rest frame of the decaying particle. When boosted, this renders the momentum of one of the fermions to be much smaller than the other one. For a longitudinal vector boson, the two fermions tend to move perpendicular to the vector boson's direction which results in more balanced momentum configurations. Therefore, the decay of a transverse vector boson is more like a QCD splitting, which makes it harder to be identified. The Higgs boson, being a spin-0 particle, has its decay products evenly distributed in its rest frame, thus the distinguishing power for a spin-0 particle is in between a transverse vector boson and a longitudinal one.  

Top jets are more complicated objects. On the one hand, compared with a $W$ jet, it is easier to distinguish a top jet from a QCD jet using kinematic information because it contains three hard subjets. One may also utilize the presence of a $b$ jet and other kinematic variables such as the helicity angle \cite{Kaplan:2008ie} to improve the top tagging efficiency. On the other hand, top quark is a colored particle, and its radiation pattern is more like a QCD jet than a boosted $W$, which makes it more difficult to use radiation variables to identify a top jet. Nonetheless, a top jet contains a $W$ jet among its decay products. One may try to use the $W$ tagging method or similar techniques to improve top tagging. This merits a detailed study. Here we only note the difference in charged multiplicity between  top jets and QCD jets. Similar to Sec.~\ref{sec:ee}, we examine top jets and QCD jets in the same fixed momentum configuration: we consider $e^+e^-\rightarrow t\bar t$ and $e^+e^-\rightarrow q\bar q gg$ events. Denoting the beam direction as the $z$ axis, we let the top move in the $x$ direction and its decay products all lie in the $y-z$ plane in the top rest frame. The momenta of the two quarks from $W$ decay are set to be of the same size. We let the $qgg$ from the $e^+e^-\rightarrow q\bar q gg$ process to have the same momentum configuration as the top decay. Then we count the numbers of charged particles in the resulting top jets and QCD jets, which are shown in Fig.~\ref{fig:nch_top}. We see a clear distinction between top jets and 3-prong QCD jets. Note that, unlike the $W$ jet case, the charge is not conserved in a top jet no matter how large the jet size is. This is because the top quark is color connected to the anti-top in the opposite hemisphere, and we expect soft particles in between them which can easily change the charge of the top jet from even to odd or vice versa.  
\begin{figure}[th!]
\begin{center}
\includegraphics[width=0.7\textwidth]{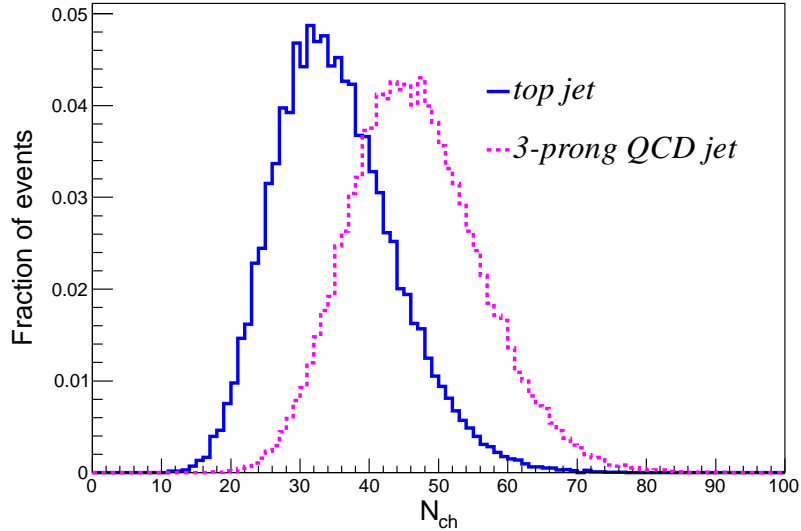} 
\caption{Charged particle multiplicities for top jets and QCD jets with $p_T=500\gev$ and fixed momentum configuration (see text) at a $e^+e^-$ machine. \label{fig:nch_top}}
\end{center}
\end{figure}

\section{Conclusion}
\label{sec:conclusion}
In this article, we have demonstrated that hadronically decaying boosted massive particles can be tagged using tracking information. Although one cannot reconstruct the masses of the decaying particle using tracks alone, the distributions of these tracks are sensitive to jet radiation patterns. In particular, the charged particle multiplicity from a color singlet particle decay is boost invariant, which serves as an excellent discriminant between $W/Z/Higgs$ jets and QCD jets. Other jet shape variables can also be calculated using charged particles alone, which is complementary to variables calculated from calorimeter information. 

We have also shown jet substructure variables can be classified to those sensitive to the jet hard splitting scale and those sensitive to the radiation pattern, which have small correlations. If two or a few variables are used to distinguish massive particle jets from QCD jets, it is most efficient to combine variables from the two categories. We have used $W$ jet tagging as an example and demonstrated that by combining the jet filtering algorithm with charged particle multiplicity or N-subjettiness, we can improve the statistical significance by a factor of $\sim 1.6$ over filtering alone. This approach simplifies the multivariate method in Ref.~\cite{wtag} that utilizes as many as 25 variables, and may find applications at the LHC especially at its early stages.

\acknowledgments
The author thanks Matthew Schwartz and Jesse Thaler for many useful discussions. The computations in this paper were run on the Odyssey cluster supported by the FAS Sciences Division Research Computing Group at Harvard University. The author is supported in part by NSF grant PHY-0804450.

\appendix

\end{document}